\documentclass[twocolumn,aps,prb,nobibnotes,superscriptaddress,bibliography]{revtex4-2}
\usepackage{amsfonts}
\usepackage{mathrsfs}
\usepackage{amsmath}
\usepackage{color}
\usepackage{graphicx}
\usepackage{bm}
\usepackage{amssymb}
\usepackage{xspace}
\usepackage{epstopdf}
\usepackage{dcolumn}
\usepackage{longtable}
\usepackage{multirow}
\usepackage{float}
\usepackage{comment}
\usepackage{braket}
\usepackage{soul}
\usepackage{multirow}
\usepackage{bbding}
\usepackage{blindtext}
\usepackage{longtable}
\usepackage{xcolor}

\usepackage{pax}
\usepackage{pdfpages}

\providecommand{\tabularnewline}{\\}

\usepackage[colorlinks=true, letterpaper=true, pdfstartview=FitV, linkcolor=blue, citecolor=blue, urlcolor=blue]{hyperref}

\makeatletter
\AtBeginDocument{\let\LS@rot\@undefined}
\makeatother

\makeatother

\begin{document}
\title{Encyclopedia of emergent particles in type-IV magnetic space groups}
\author{Zeying Zhang}
\thanks{These two authors contribute equally to this work.}
\affiliation{College of Mathematics and Physics, Beijing University of Chemical
Technology, Beijing 100029, China}
\author{Gui-Bin Liu}
\thanks{These two authors contribute equally to this work.}
\affiliation{Centre for Quantum Physics, Key Laboratory of Advanced Optoelectronic
Quantum Architecture and Measurement (MOE), School of Physics, Beijing
Institute of Technology, Beijing, 100081, China}
\affiliation{Beijing Key Lab of Nanophotonics \& Ultrafine Optoelectronic Systems,
School of Physics, Beijing Institute of Technology, Beijing, 100081,
China}
\author{Zhi-Ming Yu}
\affiliation{Centre for Quantum Physics, Key Laboratory of Advanced Optoelectronic
Quantum Architecture and Measurement (MOE), School of Physics, Beijing
Institute of Technology, Beijing, 100081, China}
\affiliation{Beijing Key Lab of Nanophotonics \& Ultrafine Optoelectronic Systems,
School of Physics, Beijing Institute of Technology, Beijing, 100081,
China}
\author{Shengyuan A. Yang }
\affiliation{Research Laboratory for Quantum Materials, Singapore University of
Technology and Design, Singapore 487372, Singapore}
\author{Yugui Yao}
\email{ygyao@bit.edu.cn}

\affiliation{Centre for Quantum Physics, Key Laboratory of Advanced Optoelectronic
Quantum Architecture and Measurement (MOE), School of Physics, Beijing
Institute of Technology, Beijing, 100081, China}
\affiliation{Beijing Key Lab of Nanophotonics \& Ultrafine Optoelectronic Systems,
School of Physics, Beijing Institute of Technology, Beijing, 100081,
China}
\begin{abstract}
The research on emergent particles in condensed matters has been attracting
tremendous interest, and recently it is extended to magnetic systems.
Here, we study  the emergent particles stabilized
by the symmetries of type-IV magnetic space groups (MSGs). Type-IV MSGs feature a special time reversal symmetry $\{\mathcal{T}|\boldsymbol{t}_0\}$, namely, the time reversal operation followed by a half lattice translation,
which significantly alters the symmetry conditions for stabilizing the
band degeneracies. In this work, based on symmetry analysis and modeling,
we present a complete classification of emergent particles in type-IV
MSGs by studying all possible (spinless and spinful, essential and
accidental) particles in each of the 517 type-IV MSGs. Particularly,
the detailed correspondence between the emergent particles and the type-IV MSGs that can host them are given in easily accessed interactive tables, where the basic information of the emergent particles, including the symmetry conditions, the effective Hamiltonian, the band dispersion and the
topological characters can be found. According to the established  encyclopedia,
we find that several emergent particles that are previously believed to exist only in spinless systems will occur in spinful systems here, and vice versa, due to the $\{{\cal T}|\boldsymbol{t}_{0}\}$ symmetry. Our work not only deepens the understanding of the symmetry conditions for realizing emergent particles but also provides specific
guidance for searching and designing materials with target particles.
\end{abstract}
\maketitle
\textit{\textcolor{blue}{Introduction}}\textit{.} In crystals, the
atoms are arranged in an orderly manner, forming crystal lattices
extending in three directions in real space and leading to periodic
band structure in momentum space \cite{bradley_mathematical_2009}. The symmetries of the bare crystal lattice are described by the space groups, where all the operators can be made unitary. By further including the spin and orbital degrees of freedom in the crystal,  anti-unitary operations such as time reversal and its combinations with certain spatial symmetries need to be considered, and the extension leads to the magnetic space groups. Correspondingly,  the energy bands of crystals should be labeled by the corepresentation of the relevant MSG \cite{belov1956color1,belov1956color2}.

With certain symmetry conditions, the energy bands may form degeneracies in the Brillouin zone (BZ), giving rise to topological semimetal states \cite{michel_connectivity_1999,burkov_topological_2011,
	wan_topological_2011,
	fang_multi-weyl_2012,
	yang_classification_2014,
	lv_experimental_2015,
	young_dirac_2015,
	huang_topological_2016,
	bradlyn_beyond_2016,
	bzdusek_nodal-chain_2016,
	yan_nodal-link_2017,
	wu_nodal_2018,
yu_quadratic_2019}.
Because the degeneracies are singularities in momentum
space, the excitations around the degeneracies show many novel
and intriguing phenomena, such as chiral anomaly, quantum vortex,
chiral Landau levels and divergent optical responses \cite{zyuzin_topological_2012,burkov_chiral_2014,huang_observation_2015,morimoto_semiclassical_2016,hirschberger_chiral_2016,liu_quantum_2017,moore_optical_2019,yu_quadratic_2019,okamura_giant_2020,yuan_discovery_2020,song_topological_2021,song_topological_2021}. Hence the topological
semimetals have been one of the most active research fields in the past ten years \cite{
	hasan_colloquium:_2010,bradlyn_topological_2017,
	armitage_weyl_2018,chiu_classification_2016,po_symmetry-based_2017}.

Rooted in the MSG symmetries and the corepresentations, the
band degeneracies take diverse forms, and each kind of degeneracy
can  exhibit distinct properties \cite{bradley_magnetic_1968,bradley_mathematical_2009,zhang_topological_2015,watanabe_structure_2018,zhang_magnetic_2021,elcoro_magnetic_2021,wang_antiferromagnetic_2017}. Thus, of fundamental importance
to list and classify all possible band degeneracies,
along with their symmetry conditions.
A complete classification of emergent particles in type-II and type-III
MSGs have recently been presented by us in Ref.~\cite{yu_encyclopedia_2021} and Ref.~\cite{liu_encyclopedia_2021},
respectively. Here, we complete the last missing piece of the project, i.e., the classification for the type-IV MSGs.

There are 517 type-IV MSGs, of which share the following structure
\begin{equation}
\text{\textbf{M}}=\text{\textbf{G}}+\{{\cal T}|\boldsymbol{t}_{0}\}\text{\textbf{G}},\label{eq:MSG4}
\end{equation}
where $\text{\textbf{G}}$ is a space group, $\mathcal{T}$ is the time reversal operation, and $\boldsymbol{t}_{0}$
is a half lattice translation. The "shifted" time reversal operation $\{\mathcal{T}|\boldsymbol{t}_0\}$ connects two lattice sites with opposite magnetic moments, indicating that type-IV MSGs describe systems with certain antiferromagnetic orders. The appearance of $\{{\cal T}|\boldsymbol{t}_{0}\}$
is also reminiscent of nonsymmorphic symmetries,
which are point group operations followed by a fractional lattice translation.
Because of $\{\mathcal{T}|\boldsymbol{t}_0\}$, although the crystals belonging to type-IV MSG are magnetic
systems without the pure ${\cal T}$ symmetry, its energy band still
exhibit the ${\cal T}$ symmetry, reflected as $E_{n}(\boldsymbol{k})=E_{n}({\cal T}\boldsymbol{k})$.
However, the $\{{\cal T}|\boldsymbol{t}_{0}\}$ here is quiet different from the pure ${\cal T}$ symmetry but similar to the nonsymmorphic spatial operators. The extra fractional
translation  
may lead new possibilities of the emergent particles in type-IV MSGs.  
For example, consider the eight time-reversal invariant momenta (TRIMs) in spinless case, $\{{\cal T}|\boldsymbol{t}_{0}\}^2= e^{-2i\boldsymbol{k}\cdot \boldsymbol{t}_0}$  will generate double degeneracy at TRIMs which satisfy $2\boldsymbol{k}\cdot \boldsymbol{t}_0=\pi$. This is distinct from other space groups, where the pure  ${\cal T}$ symmetry can lead to the Kramers double degeneracy only for spinful systems. 


\begin{table*}[t]
	\caption{The main results of the difference between type-IV magnetic space
		group and type-II MSG. \textquotedblleft$\protect\surd$" ( \textquotedblleft$\times$")
		means there (do not) exist corresponding emergent particles in magnetic
		space groups. \textquotedblleft II" and \textquotedblleft IV" represent
		the type-II MSG and type-IV magnetic space group, respectively. The
		\textquotedblleft$\protect\surd$" with red color means that the
		emergent particles exist in type-IV magnetic space group but not exist
		in type-II MSG, and the \textquotedblleft$\protect\surd$" with green
		color means that the emergent particles exist in type-II MSG but not
		exist in type-IV magnetic space group. ``Single (Double)"
		is for emergent particles  with (without) SOC. }
	\label{tab:com1}
	\begin{ruledtabular}
		\begin{tabular}{ll|cc|cc|ll|cc|cc}
			\multicolumn{1}{l}{\multirow{2}{*}{Notation}} & \multicolumn{1}{l|}{\multirow{2}{*}{Abbr.}} & \multicolumn{2}{l|}{Single} & \multicolumn{2}{l|}{Double} & \multicolumn{1}{l}{\multirow{2}{*}{Notation}} & \multicolumn{1}{l|}{\multirow{2}{*}{Abbr.}} & \multicolumn{2}{l|}{Single} & \multicolumn{2}{l}{Double}\tabularnewline
			&  & II  & IV  & II  & IV &&  & II  & IV  & II  & IV \tabularnewline
			\hline
			Charge-1 Weyl point & C-1 WP & $\surd$ & $\surd$ & $\surd$ & $\surd$ &Charge-2 Weyl point & C-2 WP & $\surd$ & $\surd$ & $\surd$ & $\surd$\tabularnewline
			Charge-3 Weyl point & C-3 WP  & $\surd$  &  $\surd$  & $\surd$  & $\surd$ & Charge-4 Weyl point & C-4 WP  & $\surd$  & $\surd$  & $\times$  & {\color{red} $\surd$}  \tabularnewline
			&&&&&&&&&&&\tabularnewline
			Triple point  & TP  & $\surd$  & $\times$  & $\surd$  & $\times$& Charge-2 triple point & C-2 TP & $\surd$ & $\surd$ & $\surd$ & $\surd$ \tabularnewline
			Quadratic triple point  & QTP & $\surd$ & $\surd$ & $\times$ & $\times$ & Quadratic contact triple point & QCTP & $\surd$ & $\surd$ & $\times$ & $\times$  \tabularnewline
			&&&&&&&&&&&\tabularnewline
			Dirac point & DP & $\surd$ & $\surd$ & $\surd$ & $\surd$&Charge-2 Dirac point & C-2 DP & $\surd$ & $\surd$ & $\surd$ & $\surd$\tabularnewline
			Charge-4 Dirac point & C-4 DP  & $\times$  & {\color{red} $\surd$}  & $\surd$  & $\surd$  &
			Quadratic Dirac point & QDP & $\surd$ & $\surd$ & $\surd$ & $\surd$ \tabularnewline
			Charge-4 quadratic Dirac point & C-4 QDP  & $\times$  & {\color{red} $\surd$}  & $\surd$  & $\surd$   &
			Quadratic contact Dirac point & QCDP & $\times$ & $\times$ & $\surd$ & $\surd$ \tabularnewline
			Cubic Dirac point & CDP & $\times$ & $\times$ & $\surd$ & $\surd$&
			Cubic crossing Dirac point & CCDP & $\surd$ & $\surd$ & $\times$ & $\times$ \tabularnewline
			&&&&&&&&&&&\tabularnewline
			Sextuple point & SP & $\surd$ & $\surd$ & $\surd$ & $\surd$ &
			Charge-4 sextuple point  & C-4 SP  & $\times$  & {\color{red} $\surd$}  & {\color{green}$\surd$}  & $\times$  \tabularnewline
			Quadratic contact sextuple point & QCSP  & $\times$  & $\times$  & {\color{green}$\surd$}  & $\times$  &
			&&&&&\tabularnewline
			&&&&&&&&&&&\tabularnewline
			Octuple point  & OP & $\times$ & $\times$ & $\surd$ & $\surd$ &&&&&& \tabularnewline
			&&&&&&&&&&&\tabularnewline
			Weyl nodal line & WNL & $\surd$ & $\surd$ & $\surd$ & $\surd$&
			Weyl nodal line (net) & WNLs & $\surd$ & $\surd$ & $\surd$ & $\surd$ \tabularnewline
			Quadratic nodal line & QNL & $\surd$ & $\surd$ & $\surd$ & $\surd$&
			Cubic nodal line  & CNL & $\times$ & $\times$ & $\surd$ & $\surd$ \tabularnewline
			&&&&&&&&&&&\tabularnewline
			Dirac nodal line  & DNL & $\surd$ & $\surd$ & $\surd$ & $\surd$ &
			Dirac nodal line (net) & DNLs & $\surd$ & $\surd$ & $\surd$ & $\surd$ \tabularnewline
			&&&&&&&&&&&\tabularnewline
			Nodal surface  & NS & $\surd$ & $\surd$ & $\surd$ & $\surd$ &
			Multiple nodal surfaces  & NSs & $\surd$ & $\surd$ & $\surd$ & $\surd$ \tabularnewline
		\end{tabular}
	\end{ruledtabular}
	
\end{table*}

In this work, we systematically study all the possible emergent
particles protected by the symmetries of type-IV MSGs, and classify
the emergent particles from four aspects, namely, the dimension of degeneracy
manifold, the degree of degeneracy, the type of dispersion, and the
topological charge. The main results are shown in Table 1, where one
can find most of the emergent particles (except QCSP ) can be realized
in type-IV MSGs. The corepresentation information and the possible
emergent particles, including spinless and spinful, essential and
accidental particles for each of the 517 type-IV MSGs are given in
easily accessible interactive tables in S5 of Supplemental Material (SM). We also compare the differences
between results of type-IV MSGs and that of type-II MSGs (nonmagnetic
systems with pure ${\cal T}$ symmetry) in table~\ref{tab:com1}. For type-II MSGs,
the C-4 WP only appears in spinless systems while the C-4 DP, the
C-4 QDP and the C-4 SP only appear in spinful systems \cite{yu_encyclopedia_2021}. In contrast,
for type-IV MSGs the C-4 SP only appear in spinless systems, and the
C-4 WP, the C-4 DP, and the C-4 QDP can appear in both spinless and
spinful systems, due to the $\{{\cal T}|\boldsymbol{t}_{0}\}$ symmetry.
We construct concrete lattice models to demonstrate the existence
of C-4 WP in spinful systems and the existence of C-4 DP in spinless
systems. Together with Ref.~\cite{yu_encyclopedia_2021} and \cite{liu_encyclopedia_2021}, our work accomplishes the grand task of classifying all possible emergent particles in periodic lattices.

\textit{\textcolor{blue}{Rationale}}\textit{.} To study the degeneracies
stabilized by the symmetries of type-IV MSGs, we should obtain the
corepresentation information of type-IV MSGs. The method to calculate
the corepresentation information of type-IV MSGs are exactly same
as that of type-II MSGs \cite{bradley_mathematical_2009}. Specifically,
the corepresentation of a type-IV MSG $\text{\textbf{M}}$ can be
induced from the small corepresentations of $M_{\boldsymbol{k}}$,
where $M_{\boldsymbol{k}}$ is the magnetic little group of $\boldsymbol{k}$
in $\text{\textbf{M}}$. This step is done by our homemade package
MSGCorep \cite{liu_msgcorep_2021}. With the calculated small corepresentation
information of $M_{\boldsymbol{k}}$, we can easily identify the possible
degeneracies protected by the symmetries of $M_{\boldsymbol{k}}$
and establish the $\boldsymbol{k}\cdot\boldsymbol{p}$ Hamiltonians
expanded around the corresponding degeneracy. Many crucial properties
of the degeneracies can be inferred from the $\boldsymbol{k}\cdot\boldsymbol{p}$
Hamiltonians, such as the dimension of degeneracy manifold, the type
of band splitting, and the topological charge. A complete list of
all possible emergent particles and their detailed classification
is presented in S4 of SM, arranged for each of the 517 type-IV MSGs.

\begin{table*}[t]
	\caption{Complex emergent particles existing in type-IV MSGs. The format of this table is similar to Tab.~\ref{tab:com1}}
	\label{tab:com2}
	\begin{ruledtabular}
		\begin{tabular}{ll|cc|cc|ll|cc|cc}
			\multicolumn{1}{l}{\multirow{2}{*}{Notation}} & \multicolumn{1}{l|}{\multirow{2}{*}{Abbr.}} & \multicolumn{2}{l|}{Single} & \multicolumn{2}{l|}{Double} & \multicolumn{1}{l}{\multirow{2}{*}{Notation}} & \multicolumn{1}{l|}{\multirow{2}{*}{Abbr.}} & \multicolumn{2}{l|}{Single} & \multicolumn{2}{l}{Double}\tabularnewline
			&  & II  & IV  & II  & IV &&  & II  & IV  & II  & IV \tabularnewline
			\hline
			Combined WNL and NS  & WNL/NS & $\surd$ & $\surd$ & $\surd$ & $\surd$ &
			Combined WNLs and NSs  & WNLs/NSs & $\surd$ & $\surd$ & $\times$ & $\times$ \tabularnewline
			Combined QNL and NS & QNL/NS  & $\times$  & {\color{red} $\surd$}  & $\times$  & {\color{red} $\surd$}   &
			Combined QNL and WNLs  & QNL/WNLs  & $\surd$  & $\surd$  & $\surd$  & $\surd$   \tabularnewline
			Combined QNLs and WNLs  & QNLs/WNLs  & {\color{green}$\surd$}  & $\times$  & $\times$  & $\times$  &
			Combined CNL and NS & CNL/NS  & $\times$  & $\times$  & $\times$  & {\color{red} $\surd$}   \tabularnewline
			Combined CNL and WNLs & CNL/WNLs  & $\times$  & $\times$  & $\times$  & {\color{red} $\surd$}
		\end{tabular}
	\end{ruledtabular}
\end{table*}

\textit{\textcolor{blue}{Compare with nonmagnetic systems.}} We also
compare the particles in type-IV MSGs with that in type-II MSGs (nonmagnetic
systems), and the results are shown in table~\ref{tab:com1}. All
kinds of accidental band degeneracies on high symmetry lines found in type-II MSGs can also be realized in type-IV MSGs. 
Hence, let's focus on the essential
band degeneracies. First, a notable feature for type-IV MSGs is that
the C-4 WP can exist in spinful systems. However, for nonmagnetic
systems, the C-4 WP can only exist in spinless materials. Besides,
the C-3 WP can appear as essential band crossing protected by the
single type-IV MSGs, but it is only allowed as accidental band
crossing on high symmetry line in type-II MSGs. Moreover, the C-4
DP, C-4 QDP and C-4 SP, emerged as spinful particles in nonmagnetic
can exist in spinless systems belonging to type-IV MSGs.

Second, there are more possibilities of the complex particles in type-IV
MSGs. Here the complex emergent particles refer to the degeneracies
that include two or multiple different kinds of particles and these
particles share one same degenerate point in BZ. For example, the
complex particle CNL/NS in MSG 184.196 includes a CNL residing along
$\Gamma A$ line and a NS located at $AHL$ plane, and the CNL and
NS are connected at joint point of $A$. However, such CNL/NS complex
particle cannot be realized in type-II MSGs. Besides, the complex particles
of QNL/NS, CNL/NS and CNL/WNLs can appear in type-IV MSGs but can
not occur at type-II MSGs, as shown in Table~\ref{tab:com2}.

\textit{\textcolor{blue}{C-4 WP and C-4 DP.}} According to previously
results, the C-4 WP and C-4 DP only exist in spinless and spinful systems  
respectively for type-II MSGs. However, the $\{{\cal T}|\boldsymbol{t}_{0}\}$ symmetry
brings new possibilities, making the realization of C-4 WP in spinful
systems and that of C-4 DP in spinless systems. According to our classification
{[}see S4C of SM{]}, we find that the C-4 WP can be realized in the magnetic
materials with SOC effect at the $R$ point of MSG 198.11, 212.62
and 213.66. Here, we use a concrete spinful lattice model with MSG
198.11 to demonstrate the existence of C-4 WP. Part of the result
in SM for MSG 198.11 are shown in table~\ref{tab:long}. In this
table, we present many important properties of MSG 198.11, including
the information of the corresponding BZ, the
corepresentation information of each high-symmetry momentum, and the
possible degeneracies protected by the MSG symmetries.

There are four points which need to be clarified, (i) for the spinful
cases with $I{\cal T}$ symmetry, all bands are at least doubly degenerate
in the whole Brillouin zone due to the Kramers degeneracy ($(I{\cal T})^{2}\equiv-1$)
and for spinless cases with $I{\cal T}$, one can always choose the
proper basis to obtain a real Hamiltonian, then the degeneracy manifold
for doubly degenerate point in those cases are always large than 0,
i.e. the degenerate point is on a nodal line or NS \cite{huang_topological_2016}.
(ii) The blue text in S5 of SM (marked as Greek letters and Latin capital letters)
are clickable. One can directly click them to get the full
form of corepresentation matrices and Hamiltonians. (iii) In order
to avoid redundant data in S7 of SM, the Hamiltonian marked as $H_{n}^{R_{i}}$
which means the expression of Hamiltonian is exactly the same as the
Hamiltonian in $R_{i}$ corepresentation of magnetic space group $n$,
despite their corepresentation matrices may different. (iv) The symmetry
operations for magnetic space groups are taken from Bradley and Cracknell's
(BC) book \cite{bradley_mathematical_2009}. However,
some magnetic space groups are mistaken in the BC book, which are not consistent with their BNS notations \cite{belov1956color1,belov1956color2}. These magnetic space groups are corrected in the package MSGCorep~\cite{liu_msgcorep_2021}, and hence all magnetic space groups we used here are consistent with those used in Bilbao Crystallographic Server or ISOTROPY \cite{aroyo_bilbao_2006,noauthor_isotropy_nodate}.


For the spinful systems with MSG 198.11, one has $\{{\cal T}|t_{0}\}^{2}=-1$
for $\Gamma$, $M$ and $\{{\cal T}|t_{0}\}^{2}=1$ for $R$,
$X$, as $t_{0}=(\frac{1}{2}\frac{1}{2}\frac{1}{2})$. Hence,
the band in $\Gamma$ and $M$ must at least be doubly degenerate,
while in $R$ and $X$ do not have such constraints. The generators
of corepresentation matrices for $R_{5}R_{6}$ of $R$ point can be
written as
\begin{equation}
\begin{split} & D(\{C_{2z}|\frac{1}{2}0\frac{1}{2}\})=-D(\{C_{2x}|\frac{1}{2}\frac{1}{2}0\})=\sigma_{0}\\
 & D(\{C_{3z}|000\})=\gamma \sigma_{19}=-\frac{1}{2}\sigma_{0}-i\frac{\sqrt{3}}{2}\sigma_{3}\\
 & D(\{{\cal T}|\frac{1}{2}\frac{1}{2}\frac{1}{2}\})=\sigma_{1}
\end{split}
\label{re5phq7eh4cy4bxxmlwn7znnfk3lk2c3}
\end{equation}
where the definition of $\sigma_{i}$ and
$\gamma$  can be found in S6 of SM. Then the $\boldsymbol{k}\cdot\boldsymbol{p}$ Hamiltonian at $R$
point can be solved from the constraint equations
\begin{equation}
H(\boldsymbol{k})=\begin{cases}
D(Q)H(R^{-1}\boldsymbol{k})D^{-1}(Q) & \text{if }Q=\{R|\boldsymbol{t}\}\\
D(Q)H^{*}(-R^{-1}\boldsymbol{k})D^{-1}(Q) & \text{if }Q=\{R{\cal T}|\boldsymbol{t}\}
\end{cases}
\end{equation}
where $D(Q)$ is the corepresentation matrices for symmetry operation
$Q$ with rotation part $R$ and translation part $\boldsymbol{t}$.
The $\boldsymbol{k}\cdot\boldsymbol{p}$ Hamiltonian for $R_{5}R_{6}$
is \cite{zhang_magnetickp_2021}
\begin{equation}
\begin{split}H= & \left(c_{1}+c_{2}k^{2}\right)\sigma_{0}+c_{5}k_{x}k_{y}k_{z}\sigma_{3}\\
 & +\left(c_{3}k_{x}^{2}+c_{4}k_{y}^{2}-c_{3}k_{z}^{2}-c_{4}k_{z}^{2}\right)\sigma_{1}\\
 & +\frac{\sqrt{3}}{3}\left(c_{3}k^{2}-3c_{3}k_{y}^{2}-c_{4}k^{2}+3c_{4}k_{x}^{2}\right)\sigma_{2}
\end{split}
\label{h3rzvc63z46wexl69ludqk7h9tozx69k}
\end{equation}
where $k^{2}=k_{x}^{2}+k_{y}^{2}+k_{z}^{2}$.  Such Hamiltonian host
(223) leading order of the band splitting and is a C-4 WP \cite{yu_encyclopedia_2021}. Moreover,
for nonmagnetic spinful systems, the C-2 TP can not appear at ${\cal T}$-invarient
point. But, as shown in Table II, the C-2 TP can appear at R point,
which again demonstrate the fact that type-IV MSG can exhibit many
intriguing phenomena that can not be realized in nonmagnetic systems.

\onecolumngrid

\begin{figure}[htbp]
	\includegraphics[width=1\columnwidth]{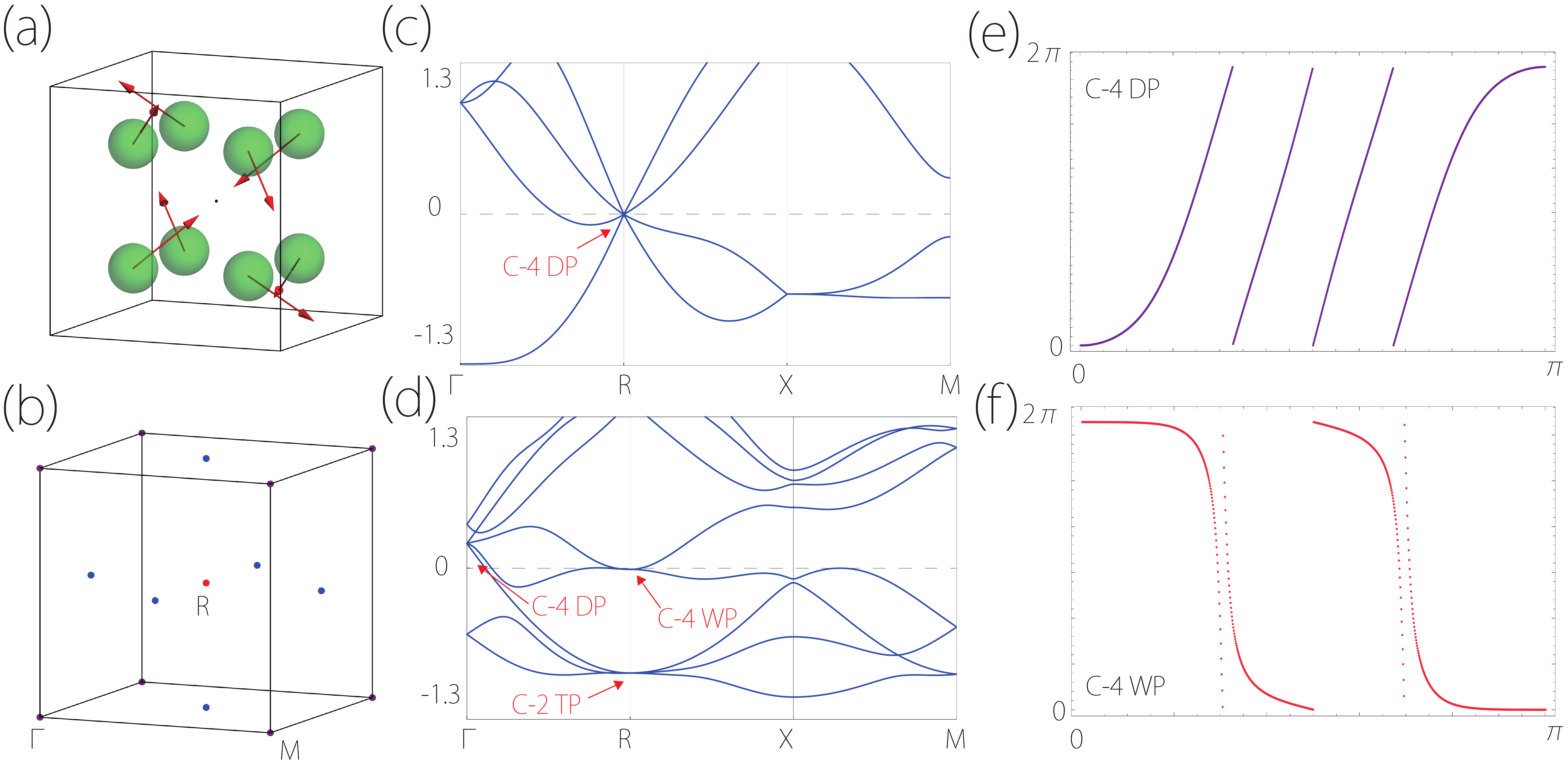} \caption{(a)Structure of lattice model. (b)Schematic showing the C-4 WP
		C-4 DP and C-1 WPs in Brillouin zone, the red, purple and blue points represent C-4 WP, C-4 DP
		and C-1 WPs, receptively. (c-d) Band structure of lattice models without SOC(c) and with SOC(d). (e-f)
		Wilson loop of C-4 DP and C-4 WP located at $\Gamma$(e) and $R$(f) and the horizontal coordinates are polar angle $\theta$ of
		the sphere.}
	\label{fig:c4wp}
\end{figure}
\twocolumngrid

To further confirm our results, we construct both spinless and spinful lattice models
for MSG 198.11. Notice the minimum Wyckoff position of 198.11 is $8a$,
consider the $\ket{\phi_{s}}$ (spinless) and $\{\ket{\phi_{s}\uparrow}$, $\ket{\phi_{s}\downarrow}\}$ (spinful)
for each atom in $8a$. Then we can construct two
tight-binding
models (see S2 of SM for Hamiltonian and parameters) \cite{zhang_magnetictb_2021}.
The structure for active atom of the lattice model is shown in Fig.~\ref{fig:c4wp}.
Moreover, for spinful case there are another C-4 DP locate at $\Gamma$ as show in Fig.~\ref{fig:c4wp}(b) The calculated band structure of this twos models are plotted in Fig.~\ref{fig:c4wp}(c-d). The topological
charge for emergent particles can be obtained by calculating
the Wilson loops on a sphere enclosing the WP (DP)
\begin{equation}
W(\theta)=\oint d\boldsymbol{k}\braket{\psi(\boldsymbol{k})|i\nabla|\psi(\boldsymbol{k})}
\end{equation}
where $\theta$ is the polar angle of the sphere. The integration
is over an enclosing loop of a constant polar angle of the sphere.
The evolution of Wilson loop for $\Gamma$ and $R$ with considering SOC  are
shown in Fig.~\ref{fig:c4wp}(e-f) which are consist with table~\ref{tab:long}.

\onecolumngrid

\LTcapwidth=\textwidth
\begin{longtable}[H]{@{\extracolsep{\fill}}*{1}{cclllllll}@{}}
	\caption{Part of the emergent particles in tpye-IV magnetic space group 198.11.
		The text above the double line of table are the basic information
		of magnetic space group, including the symbol of magnetic space group,
		the Bravais lattice, the generators of the magnetic space group, whether
		exist the combination of inversion symmetry ($I$) and ${\cal T}$,
		and whether SOC is considered. The column for $\boldsymbol{k}$ show
		the name and coordinate of high-symmetry momenta \cite{bradley_mathematical_2009}.
		The column for \textquotedblleft generators" is the rotation part
		for generators of magnetic little group of $\boldsymbol{k}$. The
		column for \textquotedblleft corep" is the information of corepresentation
		of $\boldsymbol{k}$'s magnetic little group, including the label,
		dimension and corepresentation matrices for generators. The last three
		column are the $\boldsymbol{k}\cdot\boldsymbol{p}$ Hamiltonian, type
		of emergent particles and topological charge respectively.}
	\phantomsection\label{m5q2gqx45v3jle766up336e4spcjl2ot}\\
	\multicolumn{2}{l}{\textbf{198.11}, $P_I2_13$} &
	\multicolumn{7}{r}{$\Gamma _c$, $\{ C_{2z}|\frac{1}{2}0\frac{1}{2}\},\{ C_{2x}|\frac{1}{2}\frac{1}{2}0\},\{C_{31}^{+}|000\},\{{\cal T}|\frac{1}{2}\frac{1}{2}\frac{1}{2}\}$, without $I{\cal T}$}\tabularnewline
	\hline
	\hline
	\multicolumn{2}{c}{\multirow{1}{*}{$\boldsymbol{k}$}} &
	\multicolumn{1}{c}{\multirow{2}{*}{generators}} &
	\multicolumn{3}{c}{\multirow{1}{*}{corep}} &
	\multicolumn{1}{c}{\multirow{1}{*}{$\boldsymbol{k}\cdot\boldsymbol{p}$ }} &
	\multicolumn{1}{c}{\multirow{1}{*}{node}} &
	\multicolumn{1}{c}{\multirow{2}{*}{$|{\cal C}|$}} \tabularnewline
	\cline{1-2}\cline{4-6}
	name&kinfo & &label &dim &matrices &
	\multicolumn{1}{c}{\multirow{1}{*}{Hamiltonian}}  &
	\multicolumn{1}{c}{\multirow{1}{*}{type}} & \tabularnewline
	\hline
	\multicolumn{7}{l}{Spinless emergent particles}\tabularnewline
	$\Gamma$&$000$&$ C_{2z}, C_{2x},C_{31}^{+},{\cal T}$&$\Gamma _2\Gamma _3$&2&$\hyperref[r306j8jtagrxo85m0krc4d8rp43uj3k7]{\sigma _0},\hyperref[r306j8jtagrxo85m0krc4d8rp43uj3k7]{\sigma _0},\hyperref[r32j6kttme8c7zz6f0cpodnrihez98v3]{-\gamma}  \hyperref[r84ebkjvlhmh0ycp5y0y0xp67ojqd9ax]{\sigma _{19}},\hyperref[rjte0odve4y5qn1gzrj8sywnptkfi7jv]{\sigma _1}$&$\hyperref[h3rzvc63z46wexl69ludqk7h9tozx69k]{H_{195.3}^{\Gamma _2\Gamma _3}}$&\text{C-4 WP}&$4$\tabularnewline
	&&&$\Gamma _4$&3&$-\hyperref[rso081y1uhlplk72aipq5o08b3evpo0u]{A_{37}},\hyperref[rhv3jri8e03xl8huivkzwnfn5zdtuoah]{A_{18}},\hyperref[r6wdtwfjoj8aydo6vn8ec9gbsdr1tmiu]{A_9},\hyperref[rkdcotk1mibs3cgf8rrg54vnd99kzmni]{A_0}$&$\hyperref[hj7rt7mcfbzpry3izaafn848uf60ylli]{H_{195.3}^{\Gamma _4}}$&\text{C-2 TP}&$2$\tabularnewline
	$\text{X}$&$0\frac{1}{2}0$&$ C_{2x}, C_{2y},{\cal T}$&$X_1$&2&$-\hyperref[rjte0odve4y5qn1gzrj8sywnptkfi7jv]{\sigma _1},\hyperref[rpcxr9p0jpwpu4p8jxk2gc0c5ef1wvki]{\sigma _4},\hyperref[rpcxr9p0jpwpu4p8jxk2gc0c5ef1wvki]{\sigma _4}$&$\hyperref[h255q63ftqx67qlg56g4v15ixid6rn4x]{H_{18.22}^{Y_1}}$&\text{C-1 WP}&$1$\tabularnewline
	$\text{R}$&$\frac{1}{2}\frac{1}{2}\frac{1}{2}$&$ C_{2z}, C_{2x},C_{31}^{+},{\cal T}$&$R_1$&2&$-i \hyperref[rccgu75lcqqmwgqy9kftsr3pzsk6a67b]{\sigma _3},i \hyperref[rjte0odve4y5qn1gzrj8sywnptkfi7jv]{\sigma _1},\hyperref[rthsbo9cbbk006rbi6493xp75cr2l8uf]{-\delta}  \hyperref[rmgzcxujsb3q365l5xqi5g9l2addcfx8]{\sigma _{25}},\hyperref[rpcxr9p0jpwpu4p8jxk2gc0c5ef1wvki]{\sigma _4}$&$\hyperref[h88gasn45702raovr1p79vtqt01157ku]{H_{195.3}^{\Gamma _5}}$&\text{C-1 WP}&$1$\tabularnewline
	&&&$R_2R_3$&4&$-i \hyperref[rhxl7lre1y3ptt16phjd2yuvp8dyqx2n]{\Gamma _{50}},i \hyperref[ratkp8c7n4svlnvbbb5a698s6rg0d5d9]{\Gamma _{30}},\hyperref[rt4fa0auskn31on55xbjadddkz6rwyd8]{\eta}\hyperref[r5i7u5v1ostnl6xuzb9xuf46wcr9f8rm]{\Gamma _{92}} ,-\hyperref[r57i5sp6x5bsa3zavtbnfx0m8cbs0dm5]{\Gamma _{13}}$&$\hyperref[hs7hff2dehne3g9dsz831vste7ubf5s5]{H_{195.3}^{\Gamma _6\Gamma _7}}$&\text{C-4 DP}&$4$\tabularnewline
	\multicolumn{7}{l}{Spinful emergent particles}\tabularnewline
	$\Gamma$&$000$&$ C_{2z}, C_{2x},C_{31}^{+},{\cal T}$&$\Gamma _5$&2&$-i \hyperref[rccgu75lcqqmwgqy9kftsr3pzsk6a67b]{\sigma _3},-i \hyperref[rjte0odve4y5qn1gzrj8sywnptkfi7jv]{\sigma _1},\hyperref[rbtjwmp02i4wrxinjkooxom59r8acwmr]{\delta}  \hyperref[rmgzcxujsb3q365l5xqi5g9l2addcfx8]{\sigma _{25}},\hyperref[rpcxr9p0jpwpu4p8jxk2gc0c5ef1wvki]{\sigma _4}$&$\hyperref[h88gasn45702raovr1p79vtqt01157ku]{H_{195.3}^{\Gamma _5}}$&\text{C-1 WP}&$1$\tabularnewline
	&&&$\Gamma _6\Gamma _7$&4&$-i \hyperref[rhxl7lre1y3ptt16phjd2yuvp8dyqx2n]{\Gamma _{50}},-i \hyperref[ratkp8c7n4svlnvbbb5a698s6rg0d5d9]{\Gamma _{30}},\hyperref[rrxx258w4r9d64zqlfqxrmyxim554umr]{-\eta} \hyperref[r5i7u5v1ostnl6xuzb9xuf46wcr9f8rm]{\Gamma _{92}} ,-\hyperref[r57i5sp6x5bsa3zavtbnfx0m8cbs0dm5]{\Gamma _{13}}$&$\hyperref[hs7hff2dehne3g9dsz831vste7ubf5s5]{H_{195.3}^{\Gamma _6\Gamma _7}}$&\text{C-4 DP}&$4$\tabularnewline
	$\text{M}$&$\frac{1}{2}\frac{1}{2}0$&$ C_{2x}, C_{2y},{\cal T}$&$M_5$&2&$\hyperref[rjte0odve4y5qn1gzrj8sywnptkfi7jv]{\sigma _1},-i \hyperref[rpcxr9p0jpwpu4p8jxk2gc0c5ef1wvki]{\sigma _4},\hyperref[rpcxr9p0jpwpu4p8jxk2gc0c5ef1wvki]{\sigma _4}$&$\hyperref[h255q63ftqx67qlg56g4v15ixid6rn4x]{H_{18.22}^{Y_1}}$&\text{C-1 WP}&$1$\tabularnewline
	$\text{R}$&$\frac{1}{2}\frac{1}{2}\frac{1}{2}$&$ C_{2z}, C_{2x},C_{31}^{+},{\cal T}$&$R_5R_6$&2&$\hyperref[r306j8jtagrxo85m0krc4d8rp43uj3k7]{\sigma _0},-\hyperref[r306j8jtagrxo85m0krc4d8rp43uj3k7]{\sigma _0},\hyperref[r6qpvxcywe55zluwru0as8eb8nw4enr3]{\gamma}  \hyperref[r84ebkjvlhmh0ycp5y0y0xp67ojqd9ax]{\sigma _{19}},\hyperref[rjte0odve4y5qn1gzrj8sywnptkfi7jv]{\sigma _1}$&$\hyperref[h3rzvc63z46wexl69ludqk7h9tozx69k]{H_{195.3}^{\Gamma _2\Gamma _3}}$&\text{C-4 WP}&$4$\tabularnewline
	&&&$R_7$&3&$\hyperref[rhv3jri8e03xl8huivkzwnfn5zdtuoah]{A_{18}},\hyperref[rso081y1uhlplk72aipq5o08b3evpo0u]{A_{37}},-\hyperref[r9ip8netfd8swqnulgpig3ufnrd1yya0]{A_{12}},\hyperref[rkdcotk1mibs3cgf8rrg54vnd99kzmni]{A_0}$&$\hyperref[hgicv6g5x4a2zws4yun0h1dx707e5gxj]{H_{198.11}^{R_7}}$&\text{C-2 TP}&$2$\tabularnewline
	\hline
	\hline
	\label{tab:long}
\end{longtable}

\twocolumngrid

\textit{Discussion and conclusion.} In this work, we have theoretically
listed the emergent particles in type-IV magnetic space groups. 
However, It remains an important task to identify realistic materials that can host these emergent particles. 
To put it into practice,
one can directly look up our tables in S5 of SM to find the possible emergent particles by the MSG number in the material database \cite{curtarolo_aflowlib.org:_2012,Gallego:ks5532,jain_commentary_2013}.
Besides, S4 of SM also shows in which MSGs the specific emergent particles can exist.
Experimentally, the spinful emergent particles
can be directly probed by the angle-resolved photoemission spectroscopy
(ARPES) method \cite{lv_angle-resolved_2019}. It  can also
break the ${\cal T}$ symmetry in artificial systems, which make it possible to detect spinless emergent particles in type-IV MSGs, e.g., the ${\cal T}$ breaking photonic crystals can be realized by assembling
magnetic rods \cite{wang_observation_2009, lu_symmetry-protected_2016}.
The physical properties
such as the magnetoresponse, transport behavior, topological surface
states and magneto-optical effect for emergent particles in type-IV
magnetic space group still need further investigation. For example,
the large surface density of states for magnetic higher-order nodal
lines might be beneficial for realizing surface magnetism and surface
high-temperature superconductivity \cite{zhang_magnetic_2021}.

In conclusion, with the powerful tool---group representation theory,
we establish the encyclopedia of emergent particles in type-IV magnetic
space groups which can provides useful guidance to search and study
magnetic topological materials. Two interesting directions for the
future are: (i) investigate the emergent particles in spin-space groups which have decoupled spin and lattice symmetries,
(ii) use compatibility relations to study the connectivity of energy bands and the coexistence of
emergent particles for specific Wyckoff position.

\textit{Acknowledgments.} This work is supported by
the NSF of China (Grants Nos.~12004028, 12004035, 11734003, 12061131002),
the China Postdoctoral Science Foundation (Grant No.~2020M670106),
the Strategic Priority Research Program of Chinese
Academy of Sciences (Grant No.~XDB30000000), the National Key
R\&D Program of China (Grant No.~2020YFA0308800),
, the Beijing Natural Science Foundation (Grant No.~Z190006) and the 
Singapore MOE AcRF Tier 2 (Grant No.~MOE2019-T2-1-001).

\textit{Note added.} During completion of our paper,
two independent and complementary works with a list of $\boldsymbol{k}\cdot\boldsymbol{p}$ Hamiltonians
for 1651 magnetic space groups appeared
in Ref.~\cite{tang_exhaustive_2021,jiang_kp_2021}.
{\color{red} \bf The SM of this manuscript can be found in gzipped tar source file
	(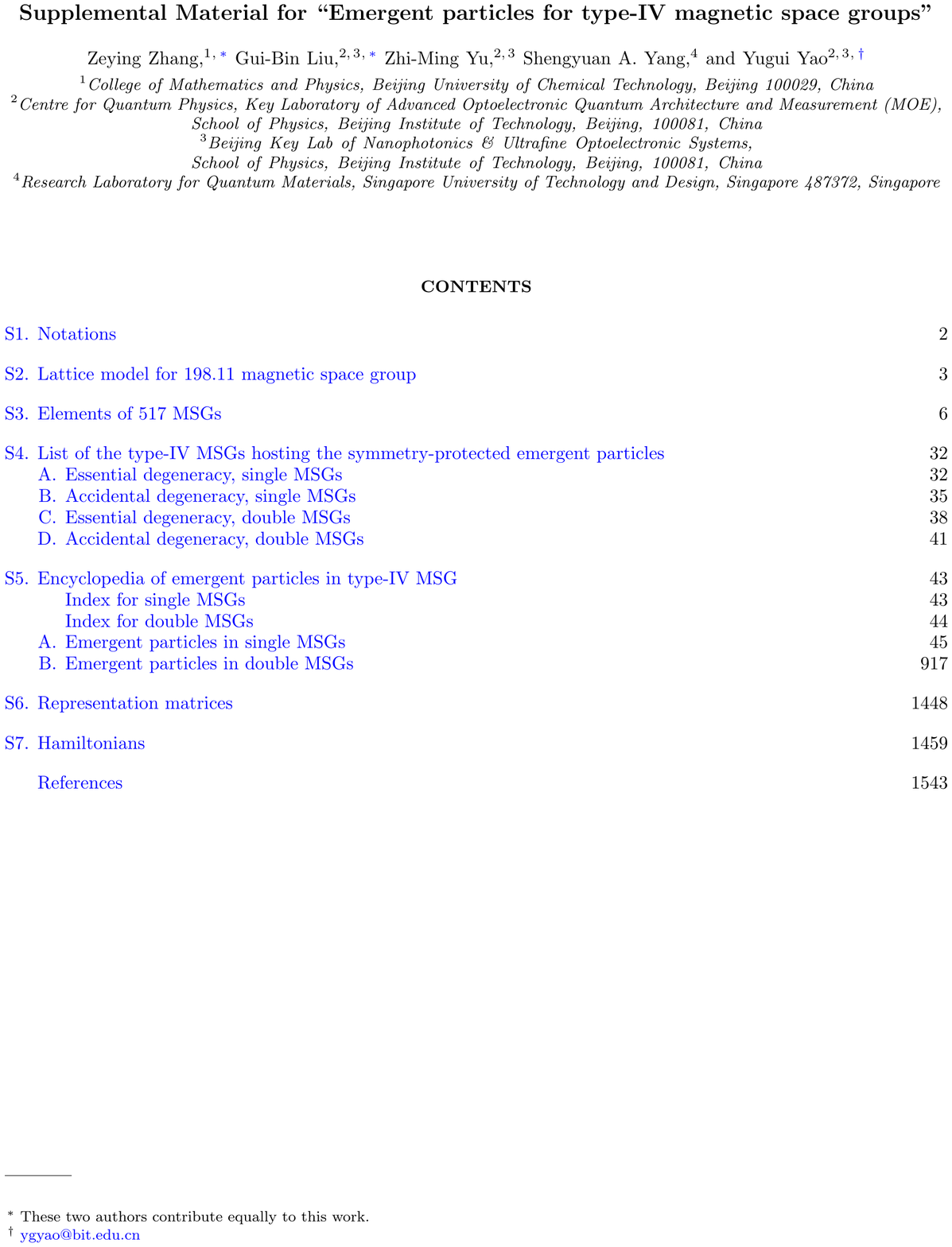).}

\bibliography{TypeIV}


\end{document}